\documentclass[two-column,showpacs,preprintnumbers,amsmath,amssymb]{revtex4}

\usepackage{subfigure}
\usepackage{times}
\usepackage{graphicx}
\usepackage{bm}
\newcommand{\bv}[1]{\mbox{\boldmath $#1$}}
\newcommand{\bvsub}[1]{\mbox{\scriptsize\boldmath $#1$}}

\def\jpbo{\emph{J. Phys. B: At. Mol. Opt. Phys.\/} }  

\def\jcp{\emph{J. Chem. Phys.\/}}
\def\pra{\emph{Phys. Rev. A\/}}

\def\prl{\emph{Phys. Rev. Lett.\/}}
\def\rpp{\emph{Rep. Prog. Phys.\/}}

\begin{document}

\preprint{APS/123-QED}

\title{Dynamic correlation in high-harmonic generation for helium atoms in intense visible  lasers}

\author{
G. A. McKenna$^1$, R. Nepstad$^{1,2}$, J. F. McCann$^1$ and D. Dundas$^3$}

\affiliation{$^1$ Department of Applied Mathematics and Theoretical Physics, School of 
Mathematics and Physics, Queen's University Belfast, Belfast BT7 1NN, UK\\
$^2$ Department of Physics and Technology, University of Bergen, N-5007 Bergen, Norway\\
$^3$ Atomistic Simulation Center, School of Mathematics and Physics, Queen's University Belfast, 
Belfast BT7 1NN, UK
}

\date{\today}
\begin{abstract}
We investigate the effect of dynamic electron correlation on high-harmonic 
generation in helium atoms using intense visible light ($\lambda =390$nm).
Two complementary approaches are used which account for correlation in an approximate manner:
 time-dependent density-functional theory and a single-active-electron model. For intensities $I \sim 10^{14}$ W/cm$^2$, the theories 
 are in remarkably good agreement for the dynamic polarization and harmonic spectrum. 
This is attributed to a low-frequency 
collective mode together with a high-frequency single-electron response due to 
the nuclear singularity, both of which dominate electron correlation effects. A time-frequency analysis is used to study 
the timing and emission 
spectrum of attosecond bursts of light. For short pulses, we find a secondary maximum below the classical cut-off. 
The imprint of the carrier-envelope phase, for the time-integrated spectral density appears at 
frequencies above the high-frequency drop-off,  consistent 
with previous studies in the infrared $\lambda \sim 800$nm.  
\end{abstract}

\maketitle

\section{\label{sec:level1}Introduction}
The development of sources of coherent ultrashort high-frequency light is a topic of great interest at present, 
with the prospect of illuminating ultrafast electron dynamics~\cite{fohlisch05, he07, ruiz06}. Over 
the past decade, high-order harmonic generation (HHG) has been used as an extreme ultra-violet 
 light source for a wide range of applications, and is a practical method for producing 
attosecond bursts from intense infrared lasers~\cite{hentschel01, paul01, balt03}. In this scheme, an intense laser is focused 
into a gas target and the high-order nonlinearity of the response of the atoms provides a 
narrow, forward-beam of high-order harmonics. The efficiency of this process depends upon the 
individual atom susceptibility and the phase matching within the gas sample.
Experiments based on this technique have led to  generation of harmonics from rare-gas atoms with 
an infrared pump laser in excess of the 300th harmonic, using very short laser pulses ($<$ 20 femtoseconds) and peak 
intensities 
in excess of $I \sim 10^{14}$ W/cm$^2$~\cite{chang, speilmann}. 

Interest is focusing on how these attosecond bursts might be controlled and optimized for applications 
in ultrafast spectroscopy. 
Availability of intense femtosecond lasers has driven research into the highly nonlinear response of atoms and molecules 
to extreme dynamic fields. The experimental tool of choice has been the Ti:Sapphire laser operating in the near infrared. 
Recently attention has turned to the use of frequency-doubled 
sources. Naturally, at very  high intensities, $I > 10^{15}$ W/cm$^2$,  double ionization is a  feature of 
the interaction. However, at intermediate intensities,  ionization is a much weaker channel and, to a good approximation, 
the process is governed by single-electron excitation, possibly leading to single ionization. However, even in this 
case,  one  would expect the electron pair to retain aspects of correlation intrinsic to the unperturbed helium atom.
The effect of correlation is a fundamental aspect of atomic structure, but its role in terms   of hyperpolarizability 
and high-harmonic generation is not fully understood. 

Dynamic correlation in helium can be treated exactly through the direct approach of solving    
the two-electron  time-dependent Schr\"odinger equation (TDSE)
\begin{equation}
i\frac{\partial}{\partial t} \Psi(\bv{r}_1, \bv{r}_2, t) = H \Psi(\bv{r}_1,  \bv{r}_2, t),
\label{SE}
\end{equation}
where $H$ is the Hamiltonian operator of the system. The solution embodies 
the dynamic polarizability that is the source of the high-frequency harmonics of interest. 
This partial-differential equation is intractable (in computational terms) for more than two electrons and 
even in the simple two-electron case, it requires several hundred processors to  obtain a numerical 
solution when infrared wavelengths are considered~\cite{parker98}. The reason for these large computational demands 
is due to the treatment of the electron-electron correlation term. In the case of stationary or metastable states of the helium 
atom, this correlation term can be accounted for, in an essentially exact manner, using basis function expansions. However, 
under transient perturbations, 
 the dynamic correlation of the atom is more problematic. Nonetheless, the problem is well known and consequently 
 this equation has received a great deal of attention, not least because of its  importance in ion-atom 
 collisions in gases and plasmas~\cite{parker07, parker06, lambropoulos05, nikolopoulos06}. 

In this paper, we investigate the role of electron correlation, by using two semi-correlated approaches. More precisely, we 
use two different models which account for correlation in an approximate manner.
 The methods employed to solve this equation are 
a time-dependent density functional theory (TDDFT) approach, recently developed to study laser-molecule 
and laser-cluster interactions~\cite{dundas:2004} and a single-active-electron (SAE) model~\cite{tong05}. In this paper, we 
study the dependence of harmonic generation from 
helium on the intensity and carrier-envelope phase (CEP), both of which have been identified as 
being important physical parameters in the determination of HHG emission properties. 

In Sec.~II below, we introduce the TDDFT approach within the exchange-only limit, and provide 
numerical details for solving the corresponding time-dependent Kohn-Sham (TDKS) equations. 
In Sec.~III, we describe the SAE method employed. In Sec.~IV, we briefly mention how the power spectrum is calculated and 
illustrate how a time-frequency analysis is performed. 
Finally we discuss and summarize the results for HHG in helium.  Atomic units are used 
throughout unless otherwise stated.

\vspace{-0.3cm}
\section{TDDFT method}
The TDDFT method has been applied extensively to the study of atomic and molecular systems
driven by external laser 
pulses~\cite{chu1,dreizler:1990,Uhlmann,dundas:2004,Calvayrac,castro:2004,otobe:2004}.
Indeed TDDFT provides one of the most detailed, practical and feasible ab initio approaches 
for tackling many-body problems. The time-dependent formulation of ground-state density 
functional theory (DFT)~\cite{hohenberg:1964,kohn:1965} was provided by Runge and 
Gross~\cite{runge:1984}, who showed that the response of a system of interacting electrons 
could be obtained from that of a set of fictitious non-interacting particles exposed to a 
time-dependent local effective potential. As well as providing a method for studying the 
time-dependent evolution of an electronic system, it also allows for the calculation of 
excited state properties of static systems. Like TDDFT, many-body effects are in principle
included exactly through an exchange-correlation functional; in practice the form of this 
functional is unknown and at best it can only be approximated.

Our implementation of TDDFT, as applied to a general spin-polarized system of $N$ electrons 
is set out in reference~\cite{dundas:2004}. In the following we describe how our approach is 
applied to the two-electron atom. The Kohn-Sham wavefunction is written as a single 
determinant of one-particle Kohn-Sham orbitals. Since helium, initially in its $^{1}S^{e}$ ground 
state, is spin degenerate then only one Kohn-Sham orbital, $\psi_{\textrm{KS}}(\bv{r}, t)$, 
is required. The electron spin 
densities are then equal, i.e.
\begin{equation}
n_{\uparrow}(\bv{r},t) = n_{\downarrow}(\bv{r},t) = 
\left|\psi_{\textrm{KS}}(\bv{r}, t) \right|^2,
\end{equation} 
where $\downarrow, \uparrow$ denotes the spin state of each electron, and so 
the total electron density is
\begin{equation}
n(\bv{r}, t) = n_{\uparrow}(\bv{r},t) + n_{\downarrow}(\bv{r},t) = 
2\left|\psi_{\textrm{KS}}(\bv{r}, t) \right|^2.
\end{equation}
The time evolution of this orbital is governed by the TDKS equation
\begin{equation}
i\frac{\partial}{\partial t} \psi_{\textrm{KS}}(\bv{r}, t) = 
H_{\textrm{KS}}\,\psi_{\textrm{KS}}(\bv{r}, t),
\label{TDKS}
\end{equation}
where
\begin{equation}
H_{\textrm{KS}}(\bv{r},  t) = -\frac{1}{2} \nabla^2 + V_{\textrm{ext}}(\bv{r},  t) + 
\int d\bv{r}^\prime \frac{n(\bv{r}^\prime,  t)}{|\bv{r}-\bv{r}^\prime|} + V_{xc}(\bv{r}, t).
\label{KSeqn}
\end{equation}
In equation~(\ref{KSeqn}) the external potential is given by
\begin{equation}
V_{\textrm{ext}}(\bv{r},  t) = V_{\textrm{ion}}(\bv{r}) + V_{\textrm{laser}}(\bv{r}, t),
\end{equation}
where $V_{\textrm{ion}}(\bv{r}, t)$ and $V_{\textrm{laser}}(\bv{r}, t)$ represent the Coulomb and 
laser-interaction terms respectively. We consider a linearly polarized laser pulse, make the 
dipole approximation and consider both the length and velocity form of the interaction. 
In a length-gauge description, the interaction term is given by
\begin{equation}
V^{(L)}_{\textrm{laser}}(\bv{r},t) = \bv{r} \cdot \hat{\bv{e}} \,E(t),
\end{equation}
where $\bv{\hat{e}}$ is the polarization direction. In the velocity gauge, the interaction term is
\begin{equation}
V^{(V)}_{\textrm{laser}}(\bv{r},t) = i A(t) \bv{\hat{e}} \,\cdot \bv{\nabla},
\end{equation}
where $A(t)$ is the vector potential defined by
\begin{equation}
A(t) =A_0 f(t) \cos (\omega_L t +\varphi),
\end{equation}
and where $\omega_L$ is the frequency, $\varphi$ the carrier-envelope phase (CEP) and $f(t)$ the  
pulse envelope given by
\begin{displaymath}
f(t) = \left\{ \begin{array}{lll}
\sin^2(\pi t/T)& { 0 \le t \le T}, \\[0.5cm]
0 & {\rm otherwise}
\end{array} \right.
\end{displaymath}
for a pulse of duration $T$.
With this form of the vector potential, the electric field is
\begin{equation}
E(t) = E_0 f(t) \sin (\omega_L t +\varphi) - \frac{E_0}{\omega_L} \frac{\partial f(t)}{\partial t} 
\cos(\omega_L t +\varphi).
\label{envelope}
\end{equation}
 The electric field amplitude ($E_0=\omega_L A_0$) is 
related to the cycle-average  intensity $I$ by $
E_0=\left( 8 \pi I/c \right)^{1/2}$,
where $c$ is the speed of light.
Similarly we define the ponderomotive energy, $U_P=E_0^2/(4\omega_L^2)$. 

The third term on the right-hand side of  equation~(\ref{KSeqn}) represents the 
Hartree potential and the fourth term incorporates the remaining exchange and correlation 
effects. This exchange-correlation potential is itself spin-degenerate for helium and can be 
written as
\begin{equation}
V_{xc}(\bv{r}, t) = V_{xc\sigma}(\bv{r}, t) = 
\left.\frac{\delta E_{xc}[n_{\uparrow}, 
n_{\downarrow}]}{\delta n_{\sigma}}\right|_{n_\sigma = n_\sigma(\bvsub{r}, t)},
\end{equation}
where $E_{xc}[n_{\uparrow}, n_{\downarrow}]$ is the exchange-correlation action. 

A crucial element in our model is the functional form of the exchange-correlation 
potential. While many sophisticated approximations to this potential have been 
developed~\cite{functionals}, the simplest is the adiabatic local density approximation in 
the exchange-only limit (xLDA). The exchange energy functional is then given by
\begin{equation}
E_x[n_{\uparrow},n_{\downarrow}] = -\frac{3}{2}\left(\frac{3}{4\pi}\right)^{1/3} 
\sum_{\sigma=\uparrow,
\downarrow} n_{\sigma}^{4/3} (\bv{r},t),
\end{equation}
from which the exchange-only potential
\begin{equation}
V_{x\sigma}(\bv{r},t) = -\left(\frac{6}{\pi}\right)^{1/3} n_{\sigma}^{1/3} (\bv{r},t),
\end{equation}
is obtained. While this approximate functional is simple to implement it does suffer from 
the drawback of containing long-range self-interaction errors:  the 
asymptotic form of the potential is exponential instead of Coulombic. The anomalous long-range 
form of the self-interaction potential means that the spectrum of single-particle
highly-excited states are incorrect. It follows that physical multiphoton resonances, or alternatively (at
 extremely high intensities) 
the tunneling process, leading to 
ionization, will be poorly represented. Nevertheless, as we will see, the gross electrical properties of the atom are 
fairly well reproduced.

Since we consider a linearly-polarized laser pulse within the dipole approximation, rotational 
symmetry around the $z$-axis is preserved at all times, and so it is appropriate to solve the TDKS equation 
using cylindrical coordinates. The electron position vector is then given by
\begin{equation}
\bv{r} = \rho \cos \phi \bv{i} + \rho \sin \phi \bv{j} + z \bv{k}.
\end{equation}
Precise numerical details of how the code 
is implemented are given in~\cite{dundas:2004}. As in~\cite{Dundas2} a finite difference 
treatment of the $z$-coordinate and a Lagrange mesh treatment of the $\rho$-coordinate 
based upon Laguerre polynomials is employed. The time-dependent Kohn-Sham equation is 
discretized in space using these grid techniques and the resulting computer code 
optimized to run on massively-parallel processors. Several parameters in the code affect 
the accuracy of the method and these are adjusted pragmatically until 
convergence is obtained. Specifically, these parameters are the number of points in the finite difference grid 
($N_z$), the finite difference grid spacing ($\Delta z$), the number of Lagrange-Laguerre 
mesh points ($N_\rho$), the scaling parameter of the Lagrange-Laguerre mesh ($h_\rho$), 
the order of the time propagator ($N_t$) and the time spacing ($\Delta t$)~\cite{dundas:2004}. 
In all the calculations presented here, converged results were obtained using the following 
parameters:  $N_z = 4485$, $\Delta z = 0.02$, 
$N_\rho = 60$, $h_\rho = 0.2027685$, $N_t = 18$ and $\Delta t = 0.01$. Incidentally, these parameters, 
without any further adjustment,  give converged (better than 1\%) TDDFT 
orbital energies for the complete first-period of elements of the periodic table.
\section{SAE method}
The single-active-electron (SAE) model~\cite{watson97, watson972, kulander92, preston96} is, perhaps,  the 
simplest and most appealing approach 
for multiphoton ionization  in which a single valence electron is released. The model  provides 
results that are cheaply produced, often to a very satisfactory degree of agreement with experiment 
and/or highly-expensive  complex many-body calculations.  As such, the model provides a useful 
benchmark in the absence of more sophisticated calculations or accurate measurements. We choose to employ spherical coordinates with the polar axis along the direction of linear 
polarization. The radius, $r$, and polar angle, $\theta$, 
are treated explicitly, whereas $\phi$ is treated analytically. Thus the field-free Hamiltonian, $H_0$, for the single-active-electron, of the helium atom initially in its 
ground-state can be written as
\begin{eqnarray}
H_0(r,\theta)& = & -{1 \over 2}\left[ {1 \over r^2} {\partial \over \partial r}  \left( r^2 {\partial \over
\partial r} \right) \right.\nonumber \\ 
&+ &\left.{1 \over r^2 \sin \theta }{\partial \over \partial \theta } \left(  \sin
\theta {\partial \over \partial \theta } \right) \right]
+V(r),
\end{eqnarray}
where the model potential $V(r)$ we chose has the form of that calculated by Tong and
Lin~\cite{tong05}
\begin{equation}
V(r) = -\frac{Z + a_1e^{-a_2 r} + a_3 r e^{-a_4 r} + a_5e^{-a_6 r}}{r},
\label{model_pot}
\end{equation}
in which,  $Z = 1.0$, $a_1 = 1.231$, $a_2 = 0.662$, $a_3 = -1.325$, $a_4 =
1.236$, $a_5 = -0.231$ and $a_6 = 0.480$. In this expression, the long-range monopole $Z=1$ is supplemented 
by short-range corrections, $\{ a_i \}$ expressing static correlation. The form of the potential  was  obtained 
from fitting to a self-interaction-free density functional. The Hamiltonian for the interacting system can be written in the form
\begin{equation}
H_{\textrm{SAE}}(r,\theta,t) = H_0(r,\theta)+ V_{\textrm{laser}}(r,\theta,t).
\end{equation}
\noindent The velocity gauge
formulation is computationally attractive since the number of partial waves
required for convergence is greatly reduced compared to the length-gauge.
This also provides us with a check on the length-gauge results.  

The
TDSE defined by equation (\ref{SE}) therefore takes the form\vspace{-0.25cm}
\begin{equation}
H_{\textrm{SAE}} (r, \theta, t) \psi_{\textrm{SAE}} (r, \theta, t)  = i \frac{\partial}{\partial t} 
\psi_{\textrm{SAE}} (r, \theta, t).
\label{eqnn}
\end{equation}
The SAE wavefunction, $\psi_{\textrm{SAE}} (r, \theta, t)$, is expanded in a direct product of 
radial and angular functions. The solid-angle normalized Legendre polynomials $Y_l^0(\theta)$ 
are efficient for the angular dimension, leading to a 
sparse interaction matrix. The radial coordinate is discretized using  $B$-spline functions 
\cite{deb78}, so that
\begin{equation}
\psi_{\textrm{SAE}}(r, \theta, t) = \sum_{n=0}^{n_{\rm max}}\sum_{l=0}^{l_{\rm max}} 
c_{nl}(t)\frac{B^k_n(r)}{r}Y_l^0(\theta), 
\label{eq:expansion}
\end{equation}
with $k$ the order of the splines, and 
where the expansion is truncated by a maximum angular momentum $l_{\rm max}$ and
the limiting number of spline functions $n_{\rm max}$. This allows us to apply Dirac's 
variation of constants method to the evolution. Taking inner products over the spatial basis functions we obtain a 
large set of ordinary, coupled 
first-order differential equations for the constants, $c_{nl}(t)$, in the form
\begin{equation}
\bv{S} \dot{\bv{c}}(t)  = -i \bv{ H }(t) \bv{c} (t),
 \label{cc}
\end{equation} 
where the overlap matrix $\bv{S}$ is due to the non-orthogonality of the
$B$-splines
\begin{equation}
S_{n'l',nl} = \delta_{l'l} \times \int_0^{r_{\rm max}} dr B^k_{n'}(r)   B^k_n(r),
 \quad |n^{\prime} -n| < k.
\end{equation}
The interaction matrix is of the form
\begin{widetext}
\begin{equation}
H_{n^{\prime} l^{\prime}, nl} = 2 \pi \int_o ^{\pi} d \theta \sin \theta \,
\int_0^{r_{\textrm{max}}} 
dr \,r  B_{n^{\prime}}^k (r) 
Y_{l^{\prime}}^0(\theta) \,H_{\textrm{SAE}} \,r^{-1} B_{n}^k (r) Y_{l}^0(\theta).
\end{equation}
\end{widetext}
\begin{figure*}
\subfigure[\small{$I$ = 1 $\times$ 10$^{14}$ W/cm$^2$}]{
\includegraphics[angle=0, width=0.45\textwidth]{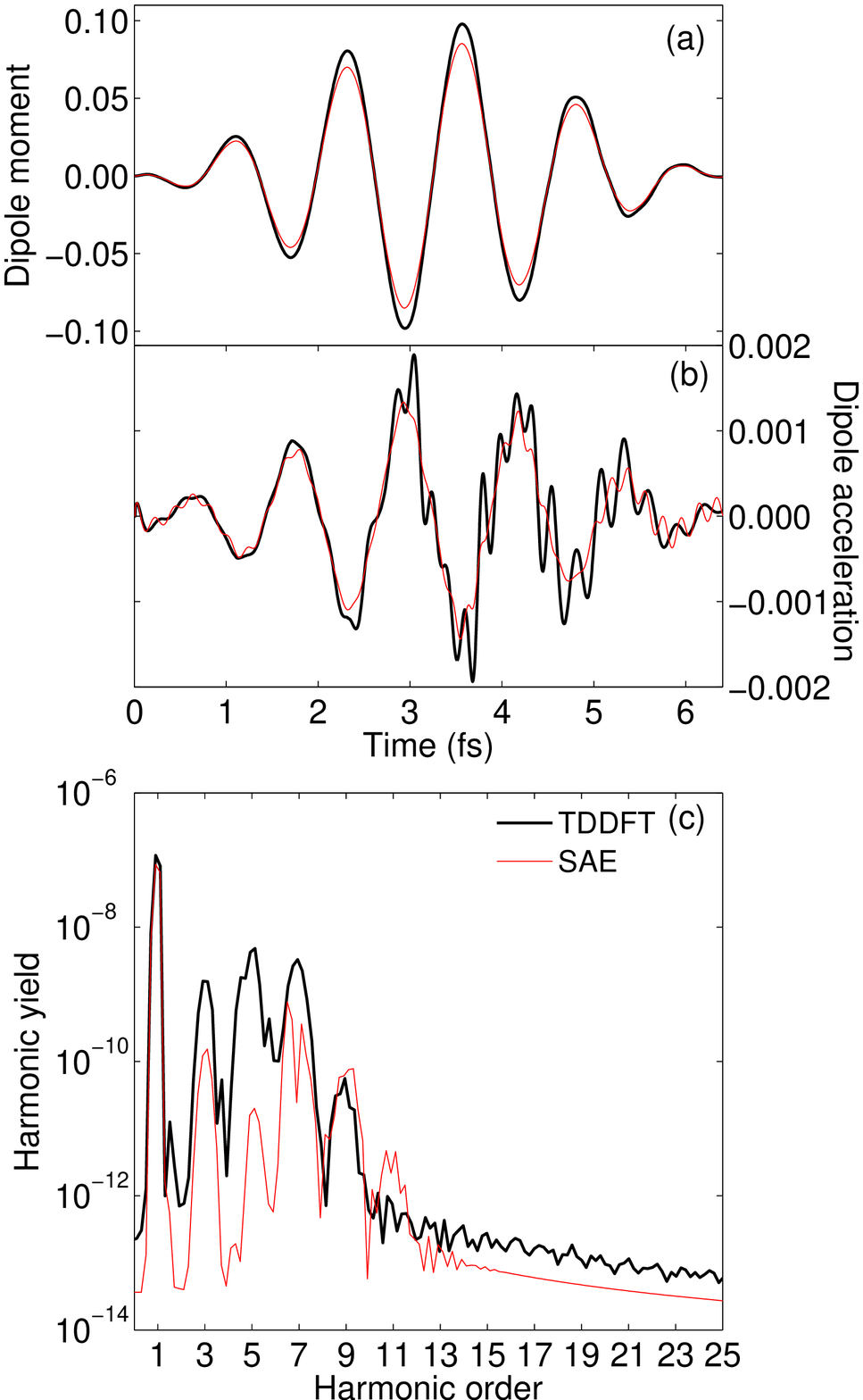}}
\subfigure[\small{$I$ = 5 $\times$ 10$^{14}$ W/cm$^2$}]{
\includegraphics[angle=0, width=0.45\textwidth]{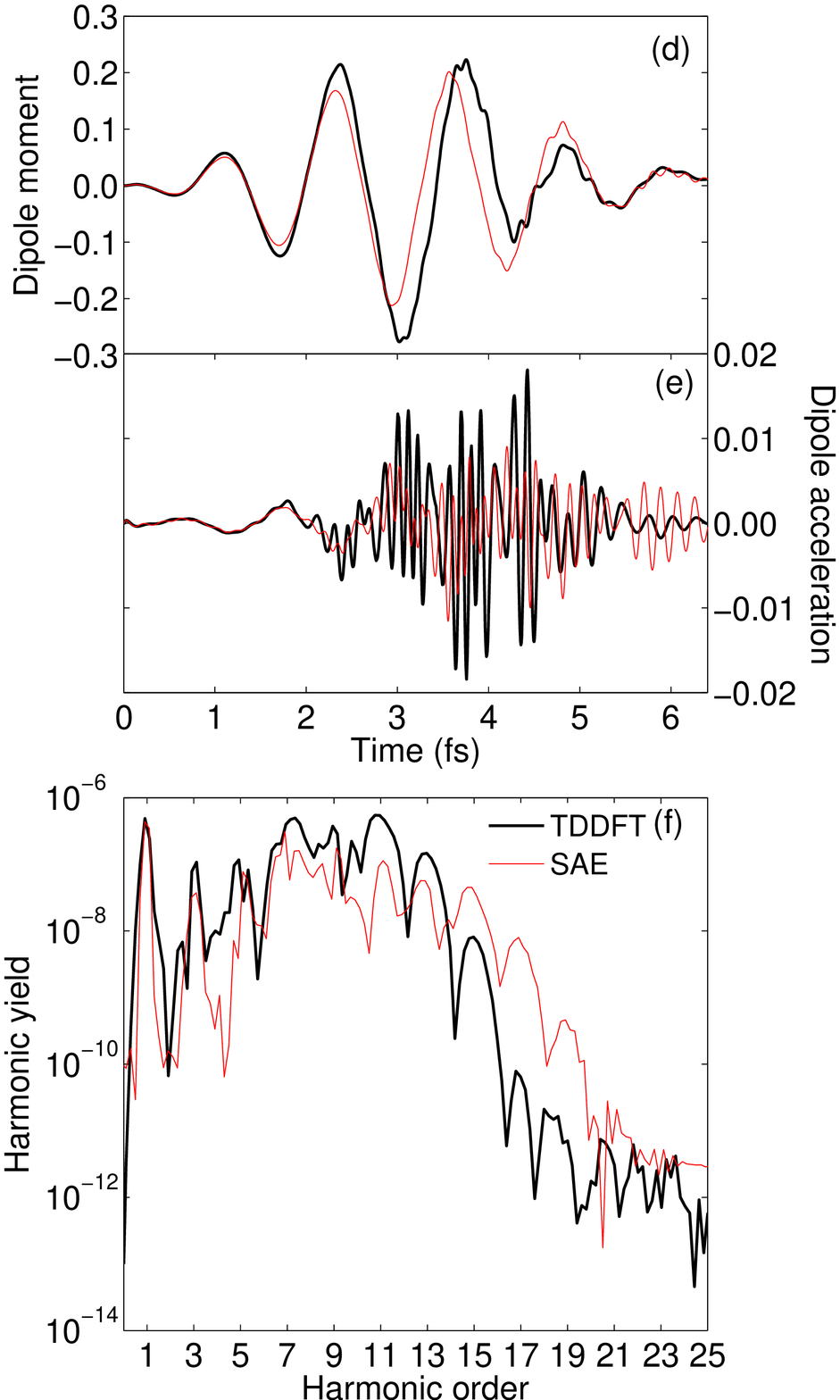}}
\caption{(Color online) Generation of high-order harmonics in a helium atom exposed
to a five cycle (6.4fs) 
linearly polarized laser pulse of wavelength $\lambda$ = 390nm for two different
models: a TDDFT model 
(black lines) and a SAE model (red lines). The plots on the left-hand-side
correspond to a laser intensity of 
$I$ = 1 $\times$ 10$^{14}$ W/cm$^2$ while the plots on the right-hand-side
correspond to a laser intensity of 
$I$ = 5 $\times$ 10$^{14}$ W/cm$^2$. Sub-plots (a) and (d) present the dipole
moment, (b) and (e) 
the dipole acceleration, and (c) and (f) the spectral density. There is remarkably
close agreement between the 
TDDFT and SAE models. The dipole (center-of-charge) motion is consistently
reproduced in phase and 
amplitude as can be seen from the dipole moment sub-plots. The irregular fluctuations 
in the dipole acceleration are also remarkably similar with a very efficient
conversion of the 7th harmonic, 
extending to the 13th harmonic at the higher intensity. The oscillations in (e) for
the SAE model after 
the pulse are due to $1s-2p$ fluorescence. There is negligible ionization for both
models at the two
intensities considered: less than 0.1\% 
for $I$ = 1 $\times$ 10$^{14}$ W/cm$^2$ and less than 2\% for $I$ = 5 $\times$
10$^{14}$ W/cm$^2$. Instead of a  classical plateau and cutoff, we have a dip and
secondary maximum below
the classical cut-off  $\omega_c$: equation (\ref{thumb}).}
\label{fig1}
\end{figure*}
The theory and applications of $B$-splines are well known~\cite{bac01,deb78},
but we briefly summarize their use for this problem. $B$-spline functions are
localized overlapping piecewise polynomials designed to reproduce the radial
oscillations of the wavepacket. These overlaps give rise to a narrow-banded
symmetric structure in $\bv{S}$, and by design, these elements can be evaluated
exactly by Gauss-Legendre quadrature. A suitable choice of spline order, $k$, is
governed by the Hamiltonian, and the radial space, $0 \leq r \leq r_{\rm max}$,
subdivided in sectors or scaled to vary the density of points, accordingly. Specifically, 
we discovered that $k=9$ order functions combine the advantages of low bandwidth and 
accurate representation of the wavefunction oscillations.

\begin{figure*}
\subfigure[\small{TDDFT}]{
\includegraphics[angle=0, width=0.45\textwidth]{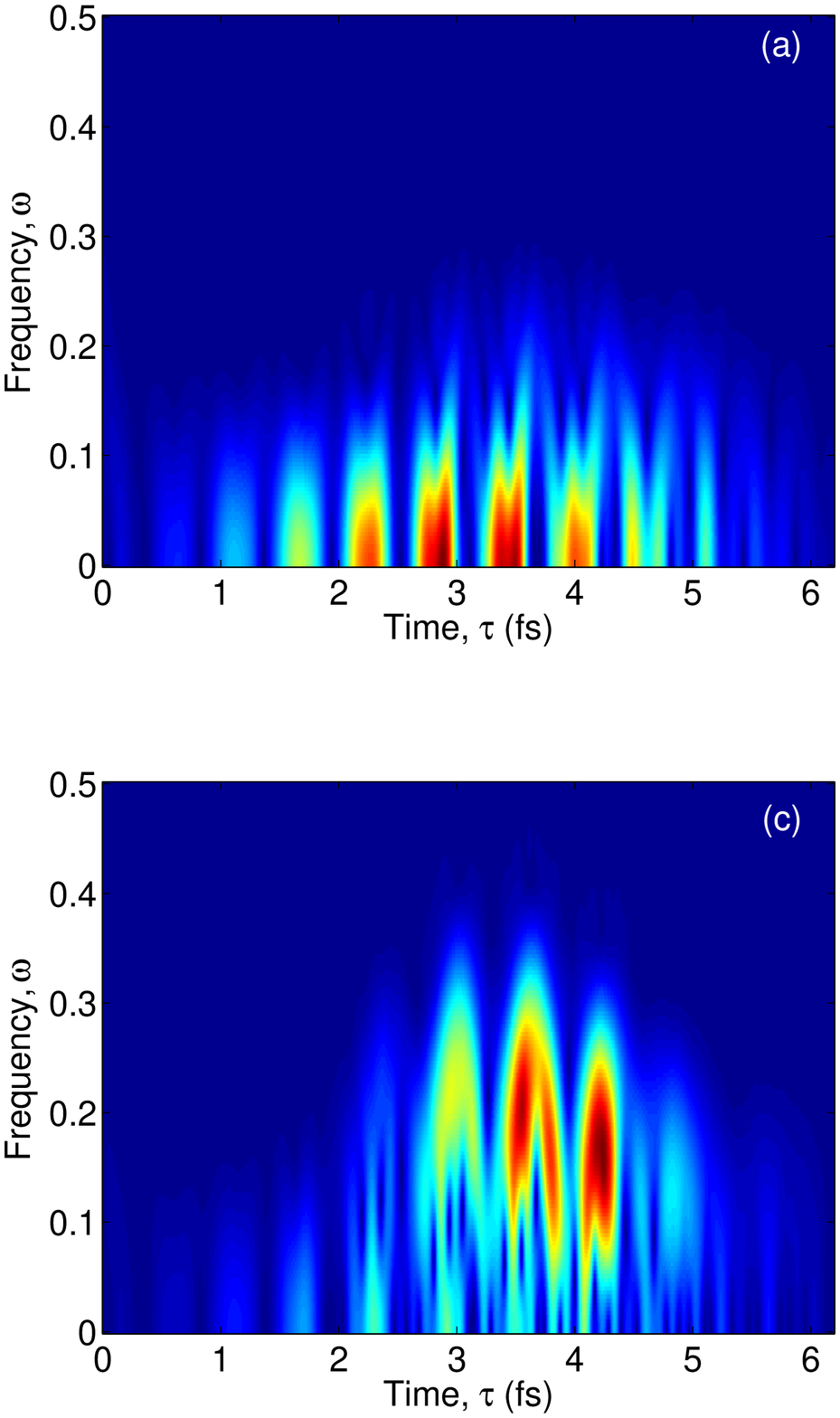}}
\subfigure[\small{SAE}]{
\includegraphics[angle=0, width=0.45\textwidth]{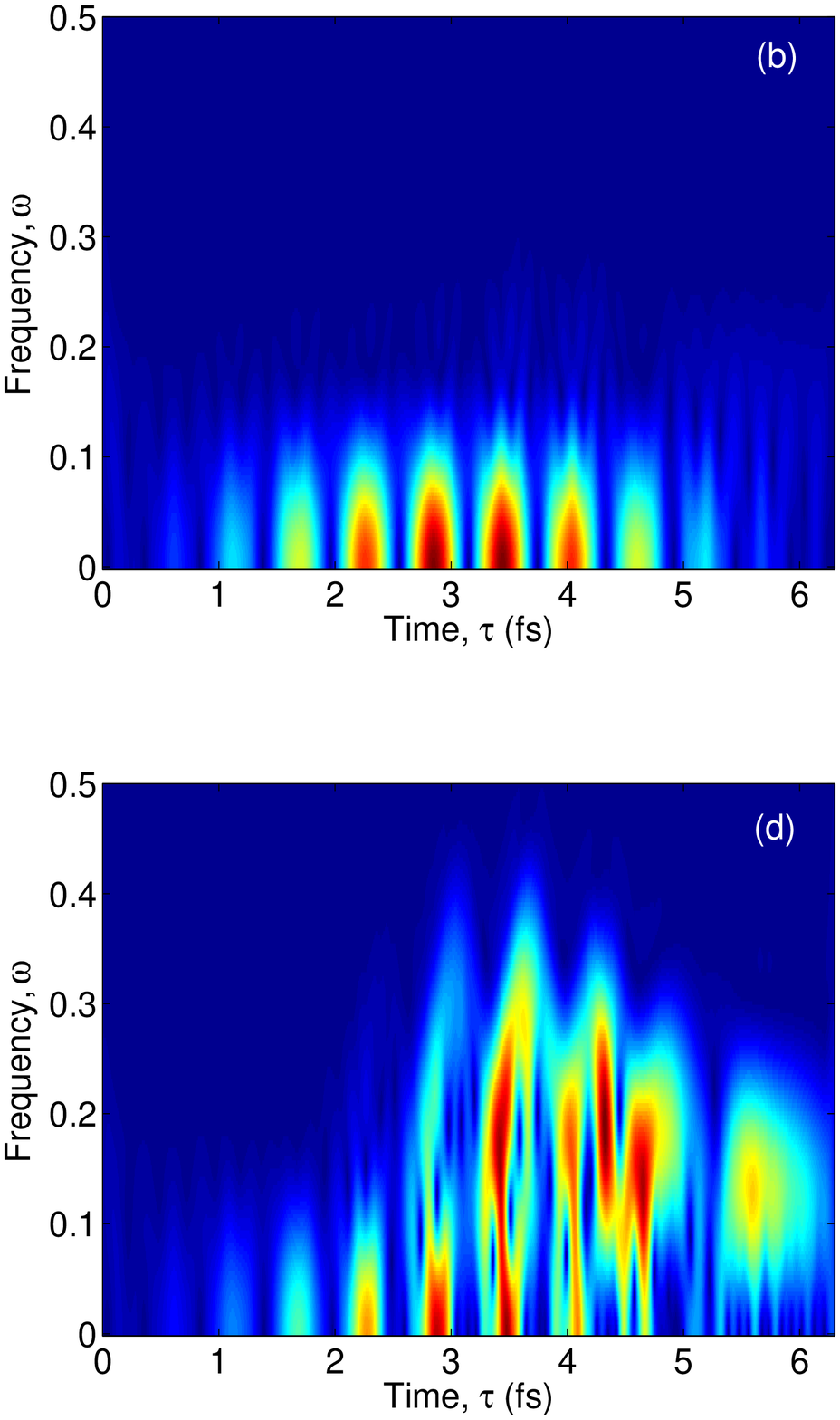}}
\caption{(Color online) Time-frequency analysis of high-order harmonic emission in a
helium atom exposed to a 
five cycle (6.4fs) linearly polarized laser pulse of wavelength $\lambda$ = 390nm
for two different models: a TDDFT model 
(left-hand plots) and a SAE model (right-hand plots). The spectral density of the STFT (equation(\ref{stftsd})) is plotted as a function of the auxiliary time $\tau$; high density is colored red and low density blue.  Two laser intensities were
considered: sub-plots (a) 
and (b) correspond to $I=1 \times 10^{14}$ W/cm$^2$ while sub-plots (c) and (d)
correspond to 
$I=5 \times 10^{14}$ W/cm$^2$. 
 }\label{fig2}
\end{figure*}

The ground-state eigenvector is calculated and normalized to provide the
stationary initial state $\bv{c}(0)$. 
Its evolution is then calculated by numerical solution of the system of equations 
(\ref{cc}). This is not a trivial task given the scale of the problem (the size of the 
interaction matrix) and the instabilities of the equations. It is well known that the 
choice of gauge has a strong effect on the  dynamical terms in the matrices. 
Not surprisingly,  this computational challenge has attracted a great deal of attention. 
While there exists  a variety of tried and tested strategies for such problems, we have 
developed our own computer codes that treat the problem in complementary ways for both
 length and velocity gauges.  This double approach  provides both a numerical validation 
 for the physical 
 results, and allows us to select the more efficient method when required. 

All calculations were performed with the {\small PYPROP} package~\cite{pyprop}, 
which is designed to provide a general interface for solution of the 
time-dependent Schr\"odinger equation. Discretization schemes and propagation 
methods are implemented following a standard interface, which enables users to 
test different methods without requiring detailed knowledge about the 
implementation. {\small PYPROP}  is written in Python,  while all the 
computationally-intensive routines have been developed in C++, 
using the blitz++ array library~\cite{blitz}. Depending on the problem at hand, the size 
of the grid and number of employed grid points will vary. 
For the range of angular momenta, we found that $l_{\rm max} = 20$ was more than 
sufficient for convergence in all cases considered here. In addition we found that 1050 B-splines were sufficient for convergence. 
Time propagation was carried out using the
Arnoldi method~\cite{par86}.

\begin{figure*}
\subfigure[\small{TDDFT: $I$ = 1 $\times$ 10$^{14}$ W/cm$^2$}]{
\includegraphics[angle=0, width=0.45\textwidth]{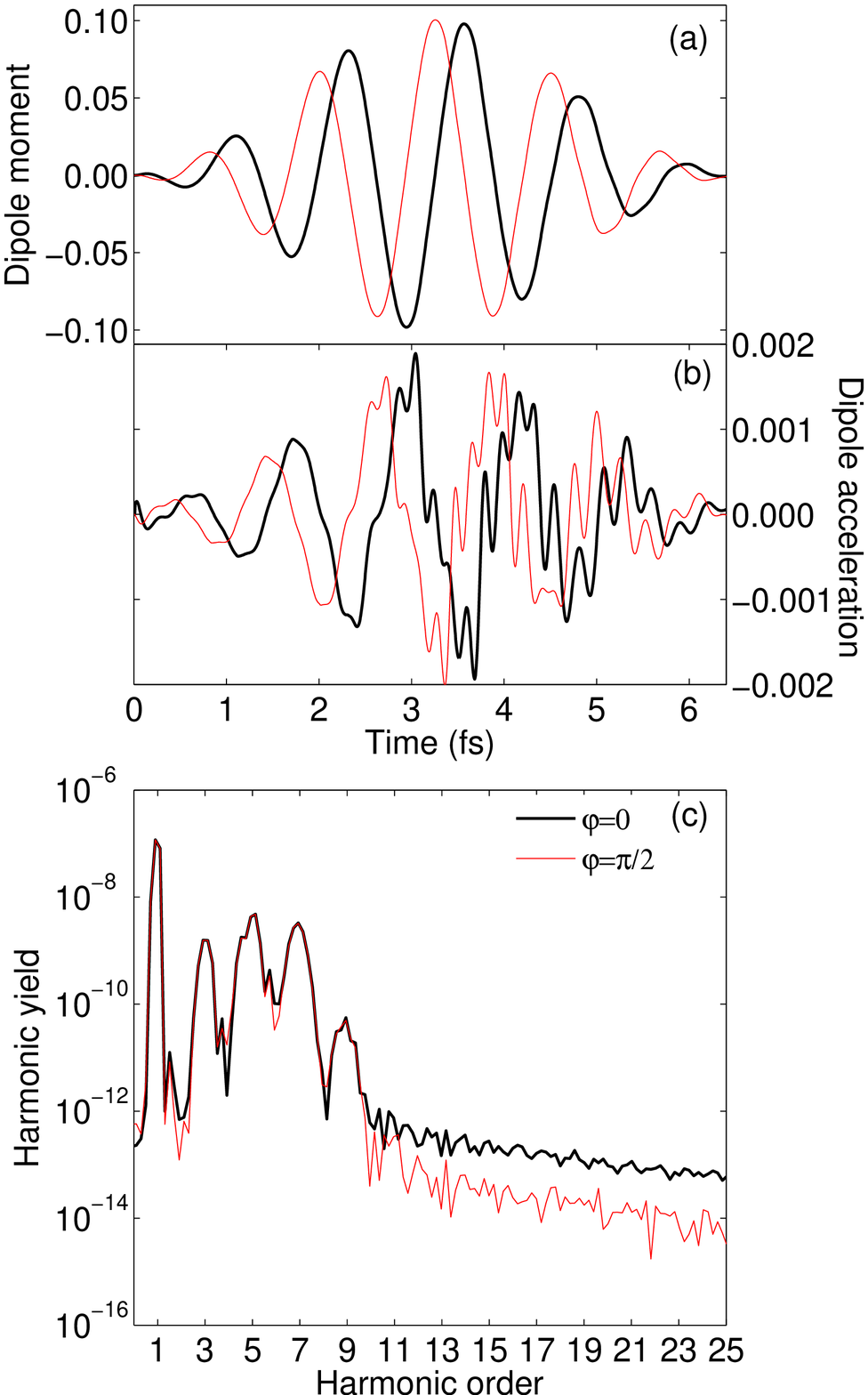}}
\subfigure[\small{SAE: $I$ = 1 $\times$ 10$^{14}$ W/cm$^2$}]{
\includegraphics[angle=0, width=0.45\textwidth]{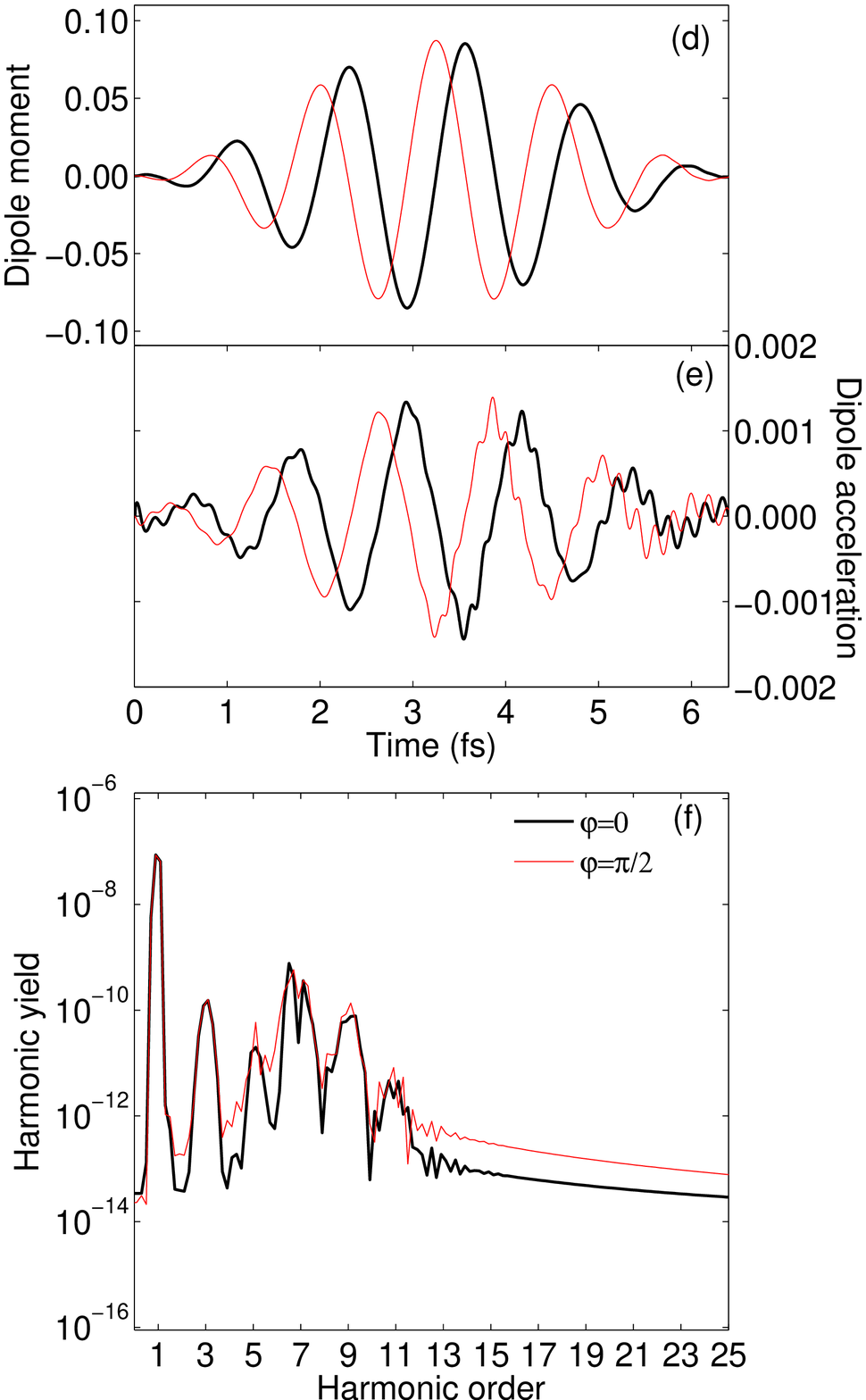}}
\caption{(Color online)  The effect of carrier-envelope phase (CEP), $\varphi$, on
harmonic generation 
in a helium atom exposed to a five cycle (6.4fs) linearly polarized laser pulse of
wavelength $\lambda$ = 390nm 
and intensity $I$ = 1 $\times$ 10$^{14}$ W/cm$^2$ for two different models: a TDDFT
model 
(left-hand plots) and a SAE model (right-hand plots). A pulse envelope defined by
equation (\ref{envelope}) 
was used and two CEP values were considered: $\varphi=0$ (black lines) and 
$\varphi=\pi/2$ (red lines). Sub-plots (a) and (d) present the dipole moment, (b)
and (e) 
the dipole acceleration, and (c) and (f) the spectral density. 
We note that, for both models, there is very little effect of CEP in the
time-integrated spectral density.}\label{fig3}
\end{figure*}

\section{Frequency Analysis}
\indent   The dipole moment is defined as
\begin{eqnarray}
d(t) &=& - \int  n(\bv{r},t) \,z \, d^3 \bv{r},\nonumber \\
&=& -2 \langle \psi(t)|z|\psi(t)\rangle.
\end{eqnarray}
According to Larmor's formula, the radiated power is proportional to the square of the  dipole acceleration. 
This can be obtained using Ehrenfest's theorem as~\cite{burnett92}
\begin{equation}
\ddot{d}(t) = - 2\langle \psi(t) |[H,[H,z]]|\psi(t) \rangle.
\label{acccusp}
\end{equation} 
The spectral density is then obtained as follows
\begin{equation}
S(\omega) =\left| \int_0^T \ddot{d}(t) e^{i \omega t} dt
\right|^2.
\end{equation}

\begin{figure*}
\subfigure[\small{TDDFT: $I$ = 5 $\times$ 10$^{14}$ W/cm$^2$}]{
\includegraphics[angle=0, width=0.45\textwidth]{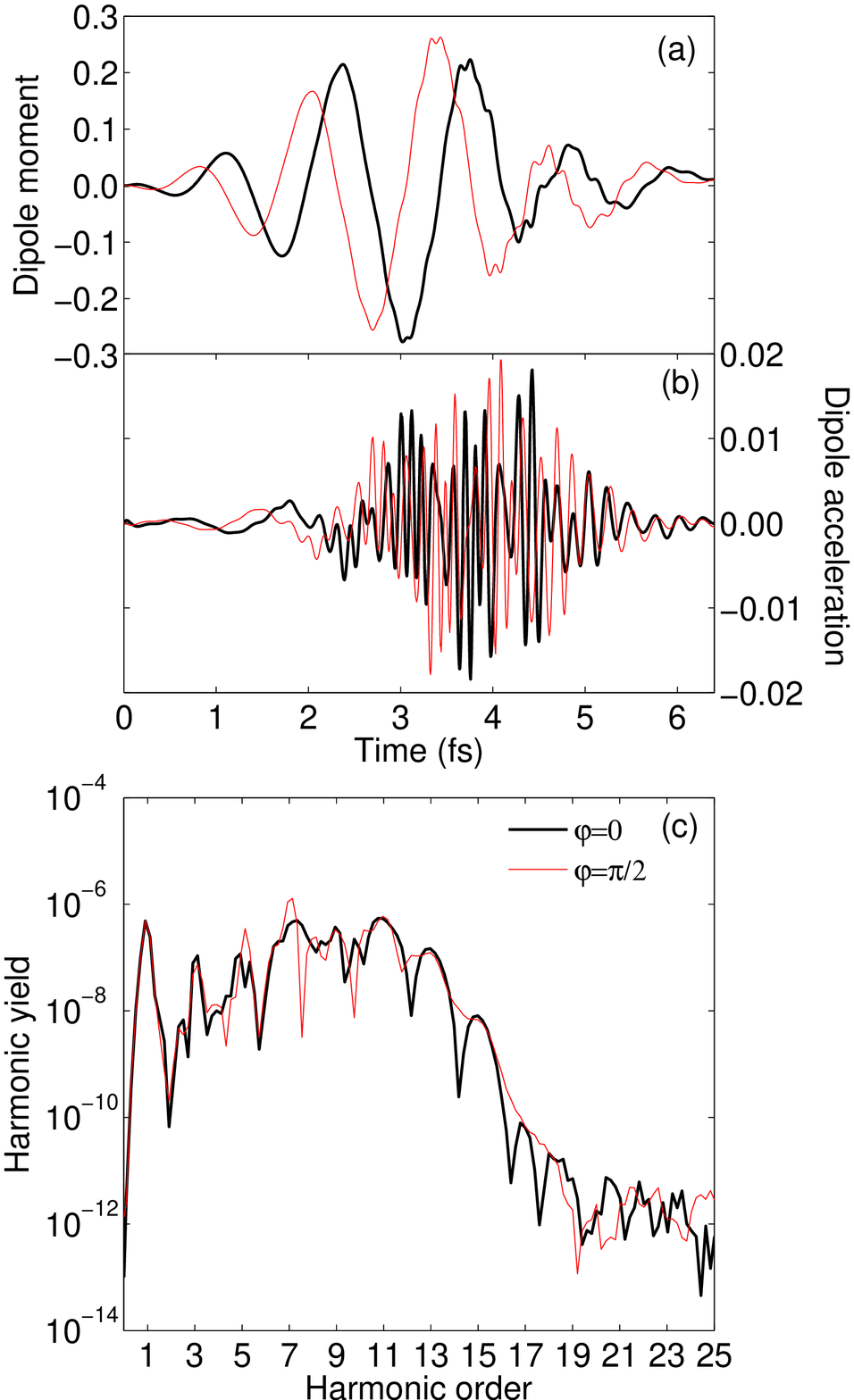}}
\subfigure[\small{SAE: $I$ = 5 $\times$ 10$^{14}$ W/cm$^2$}]{
\includegraphics[angle=0, width=0.45\textwidth]{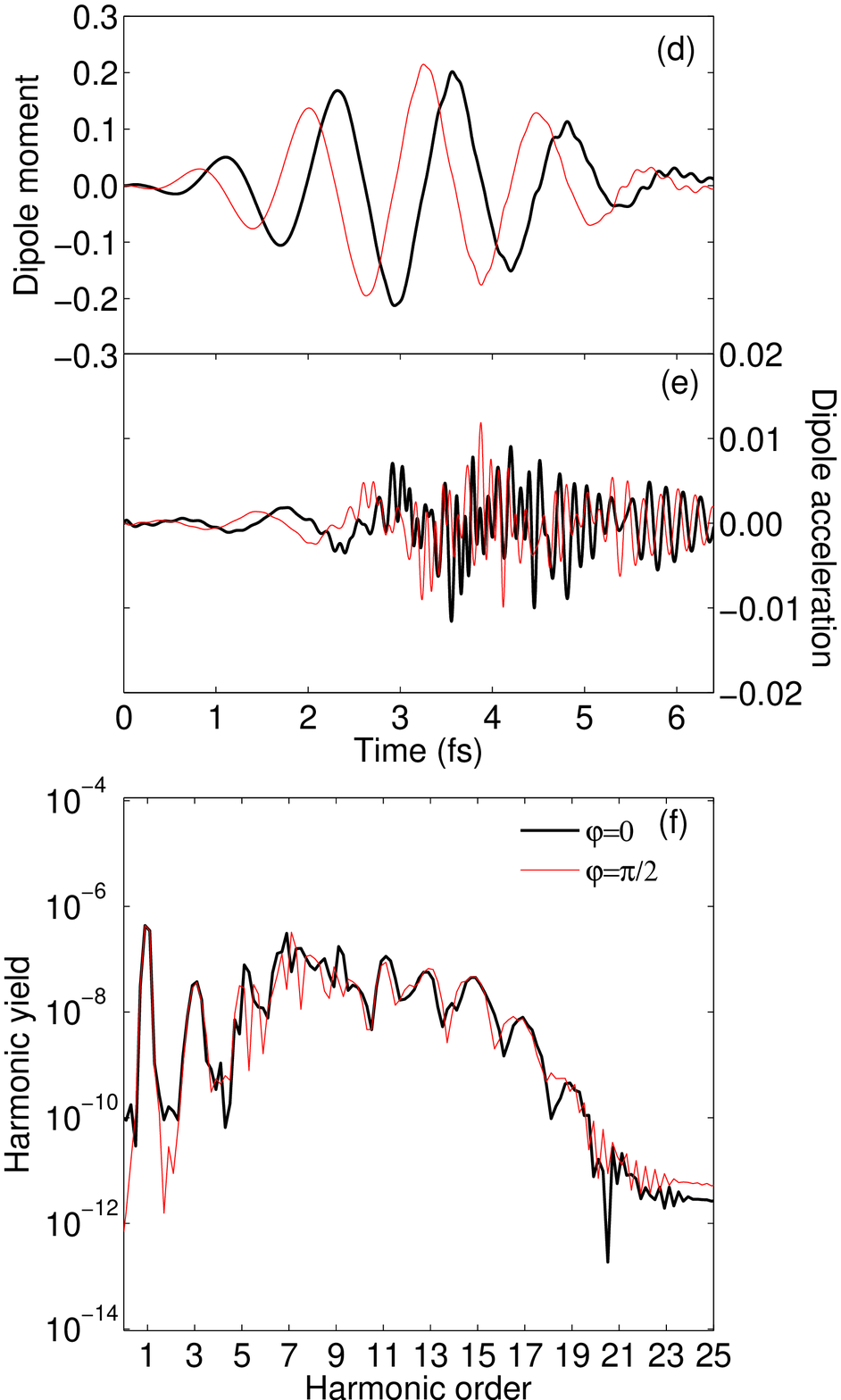}}
\caption{(Color online)  The effect of carrier-envelope phase (CEP), $\varphi$, on
harmonic generation 
in a helium atom exposed to a five cycle (6.4fs) linearly polarized laser pulse of
wavelength $\lambda$ = 390nm 
and intensity $I$ = 5 $\times$ 10$^{14}$ W/cm$^2$ for two different models: a TDDFT
model 
(left-hand plots) and a SAE model (right-hand plots). A pulse envelope defined by
equation (\ref{envelope}) 
was used and two CEP values were considered: $\varphi=0$ (black lines) and 
$\varphi=\pi/2$ (red lines). Sub-plots (a) and (d) present the dipole moment, (b)
and (e) 
the dipole acceleration, and (c) and (f) the spectral density. 
 The spectral density obtained from the TDDFT model exhibits a strong dependence on
carrier phase, both below and 
 above the secondary maximum. This behavior is consistent with a fully-correlated
numerical 
 simulation for helium in the infrared \cite{balt03}, Conversely, 
 the spectral density obtained from the SAE model lacks this feature.
}
\label{fig4}
\end{figure*}

To analyze the frequency content of a signal over time, both time and frequency information 
is needed simultaneously, i.e.~a time-frequency analysis provides valuable insight. Here 
we use the  Short-Time Fourier Transform (STFT), 
which has been found to recover dominant frequencies within a signal with reliable
accuracy~\cite{tang00}. The spectral density of the STFT is given by~\cite{stft94}
\begin{equation}
F(\omega,\tau) =  \left| \int_{0}^{+\infty } \ddot{d}(t) \,
h(t-\tau) e^{i \omega t} dt \right|^2,
\label{stftsd}
\end{equation}
where $\omega$ and $\tau$ represent the analyzing frequency and time-shift respectively, and we chose to employ a
 Hann window function given by~\cite{hann89}
\begin{equation}
h(t) =\frac{1}{2} \left[1-\cos\left(\frac{2 \pi t}{T_W} \right) \right],
\end{equation}
where $T_W$ is the window length. In our case, given the timing of the radiation bursts, 
 a natural choice for $T_W$ was the optical  period. 
\section{Results and Discussion}
\indent We present results for the dipole moment, dipole acceleration and
corresponding power spectra for the two models at two different laser intensities in Fig~\ref{fig1}. In all simulations 
a laser wavelength of $\lambda$ = 390nm was considered, corresponding to frequency-doubled Ti:sapphire wavelength. 
The two laser intensities considered are $I$ = 1 
$\times$ 10$^{14}$ and 5 $\times$ 10$^{14}$ W/cm$^2$. 
For the TDDFT method, the ground-state energy,
calculated by propagating in imaginary time, was found to be 2.7240 a.u. The 
ground-state of singly ionized helium is 1.8068 a.u., and thus the ionisation energy in the TDDFT model is
$I_p$ = 0.9172 a.u. In the SAE model $I_p$ = 0.9034 a.u. which coincides with the ionization energy of the TDDFT model. 
The good agreement 
is perhaps surprising given the very different forms of the Hamiltonian and their respective treatment 
of correlation. We also calculated the static polarizability $\alpha$, which describes 
 the quadratic (lowest-order non-zero) Stark shift.   For the TDDFT model presented  we
find $\alpha=1.76$, which is in good agreement with experiment $\alpha=1.38$ and other TDDFT calculations~\cite{banerjee:1999}. 

Let us now consider the center-of-charge oscillations in Figs.~\ref{fig1}(a) and \ref{fig1}(d). 
At both intensities,  the motion is dominated by the fundamental dipole motion. This manifests itself in the agreement of 
the fundamental frequency in the harmonic spectra for both models. This collective motion, 
in a many-body system, is expected of a tightly-bound oscillator under a low-frequency external field 
($\lambda=390$nm corresponds to $\omega_L=0.1168$ a.u.). In general, the center-of-charge motion is weakly 
dependent  on the internal forces, and would not be expected to show dynamic correlation effects. Thus the 
different models of correlation in  TDDFT and SAE models are not significant when the fundamental mode is 
observed.


Deviations between TDDFT and SAE are apparent with high-frequency quivering at the peak of the external field in
Figs.~\ref{fig1}(a) and \ref{fig1}(d), and are revealed when the dipole mode is removed. The small amplitude 
oscillations are most pronounced at the  peak of the electric field and occur near the nucleus where the 
restoring force is strongest and the electron acceleration is most intense. The same trend carries over 
to the dipole accelerations in Figs.~\ref{fig1}(b) and \ref{fig1}(d), resulting in a more intense spectral density as shown in 
Figs.~\ref{fig1}(c) and \ref{fig1}(f). 

Three features typical of HHG are~\cite{corkum93, lewenstein94, corkum07}: (i) low order harmonics, which arise from transitions from bound 
excited states to the ground-state, show a rapid decrease in intensity as expected from a perturbative 
process; (ii) a (short) plateau region, caused by transitions from the continuum, of relatively 
constant intensities; and (iii) an abrupt cut-off, at a frequency
\begin{equation}
\omega_c \sim I_P+3.2 U_P,
\label{thumb}
\end{equation}
where $I_P=0.9040$ a.u. is the ionization potential. Rather than a plateau, we observe  a dip followed by 
a secondary maximum. As the laser intensity is increased,  this secondary maximum extends to higher frequency 
and broadens. Comparison with the classical cut-off formula, equation (\ref{thumb}), for $I$ = 1 $\times$ 10$^{14}$ and
$I=5 \times 10^{14}$ W/cm$^2$, gives $\omega_c \sim  9 \omega_L$ and $\omega_c \sim 15 \omega_L $, respectively. 
Fig.~\ref{fig1} illustrates, that both the TDDFT and SAE calculations falls short of these harmonics, with 
secondary maxima nearer  the 7th and 11th harmonics.  On further investigation, we found this attenuation 
is not a modulation feature due to the pulse shape, in which the intensity variation is rapidly varying. 
A much longer pulse  ($T=25.6$fs) gives similar results for the drop-off frequency. 

Given that TDDFT and SAE  provide conflicting descriptions of the asymptotic effective potential, it is 
clear that long-range correlations in the two models will differ and the ionization rates, for example, will 
disagree. At both these laser intensities we found that ionization at the end of the pulse was negligible in the two 
models ($<$ 2\%). Moreover, in the case of short-range behavior, the correlation terms in each model will be 
negligible in comparison with the Coulomb singularity that gives rise to the Kato cusp in the wavefunction. 
Essentially correlation effects are dominated by the nuclear singularity in this region. According to equation (\ref{acccusp}) the highly-singular 
acceleration operator is strongly peaked in this region and thus the effective Hamiltonians are single-electron 
(uncorrelated hydrogenic) expressions at the highest frequency. Thus the TDDFT and SAE models share the same 
essential features in the lowest and highest frequency emission. At low frequency we have a collective 
low-frequency mode of the orbital, while the highest frequency emission is essentially single-electron 
hydrogenic behavior. So, at the extremes of the spectrum this again dominates correlation effects, while in the intermediate 
region (Fig.~\ref{fig1}(c), harmonics 3, 5 and 7) correlation corrections are significant. 

At  much higher intensity, $I=5 \times 10^{14}$ W/cm$^2$,  the laser field, $E_0 \approx 0.12$, competes 
with the binding potential and the correlation energy.  The degree of atomic excitation is greatly increased 
and we would expect significant disparity between the models. The SAE potential is weaker than the equivalent 
TDDFT effective potential.  Nonetheless, ionization is less than a few per cent and Fig.~\ref{fig1}(d) 
indicates that the center-of-charge motion is more or less the same. The dipole acceleration in 
Fig.~\ref{fig1}(e) is now smeared across the duration of the pulse, though still indicating prominent bursts 
at the half-cycle  turning points  of the moment. The time-integrated spectral density in 
Fig.~\ref{fig1}(f) shows the matching of the fundamental in both models, as was observed at the lower intensity, and again a stronger 
yield of lower harmonics in the TDDFT model. The secondary maximum is as strong as the fundamental 
peak. For example with a five-fold increase in intensity, the 9th harmonic is over 100 times more intense, as shown in
Figs.~\ref{fig1}(c) and \ref{fig1}(f). The long slow decay of the SAE spectrum in the drop-off is consistent with a 
lower ionization threshold for this model.

The time-integrated spectral densities do not reveal the attosecond dynamics of the correlation.
In Fig.~\ref{fig2}, we  present an analysis of each burst of harmonics through the STFT. The 5 cycle pulse 
provides an interaction time of  $T=6.4$fs. At the lower intensity, Fig.~\ref{fig2}(a), the burst created at 2.25fs 
is correlated to the turning point in Fig.~\ref{fig1}(a). The subsequent burst, half a cycle later at 2.9fs shows a 
bifurcation at the higher frequencies which coincides with the double-peak in the dipole acceleration in 
Fig.~\ref{fig1}(b). The irregularities in Fig.~\ref{fig1}(e) at the higher intensity are reflected in fringes within the 
main bursts in Figs.~\ref{fig2}(c) and \ref{fig2}(d) above. The peak emission time is also broadened and slightly delayed. 
In Fig.~\ref{fig2}(d), the SAE results show an extended fluorescence signal due to the $2p-1s$ emission after the 
passage of the pulse. At the higher intensity,  Fig.~\ref{fig2}(c) we notice the enhancement in high-frequencies at the 
expense of the lower frequencies, in accordance with Fig.~\ref{fig1}(c). However, the timing of the peak 
bursts in Fig.~\ref{fig2}(c) are slightly delayed compared with Fig.~\ref{fig2}(a), for example, with the TDDFT model. This delay is not 
observed for the SAE model. 

Baltuska  \textit{et al} \cite{balt03} found that HHG was very sensitive to the carrier-envelope 
phase. Their experiments with helium, using the fundamental Ti:Sapphire mode in the infrared 
$\lambda=800$nm, showed that, provided the CEP was chosen to coincide with the peak of the  envelope, 
the spectrum generated in the cutoff region loses the odd-order harmonics. These results were 
supported by theoretical simulations in the same letter. We examined whether this holds true for our two 
models at the frequency-doubled laser wavelength: $\lambda=390$nm. The results are presented in 
Figs.~\ref{fig3} and \ref{fig4}. Firstly, referring to Fig.~\ref{fig3}, the time-integrated spectral 
density, subfigures (c) and (f), show very little sensitivity to the phase $\varphi$. 
 
At the higher intensity the sensitivity to CEP is apparent in the TDDFT method but not the SAE results.
The TDDFT simulations, Fig.~\ref{fig4}(c), are consistent with experiment and simulations 
(fully-dimensional, fully-correlated) reported for $\lambda \sim 800$nm \cite{balt03}. Beyond  
the cutoff it is clearly evident that the spectrum is continuous when $\varphi =\pi/2$ and harmonic 
when $\varphi=0$. On the other hand, the SAE simulations  do not exhibit this behavior. Furthermore, 
the highest harmonic produced by the SAE model at $I = 1 \times 10^{14}$ W/cm$^2$ is 13th-order and at
$I= 5\times 10^{14}$ W/cm$^2$ is 23rd-order, while the TDDFT results extend well beyond this range. On this basis, 
it suggests that, for these parameters, a more complete model of correlation is provided by the TDDFT model. 
Of course, confirmation of this finding from a full-dimensional  two-electron simulation and experimental 
measurements would be important. 

\section{Conclusions}
In this paper, we have investigated HHG in helium subjected to short intense laser pulses, using two models of correlation: a 
TDDFT approach and a SAE model. We presented harmonic yields for two laser intensities ($I$ = 1 $\times$ 10$^{14}$ W/cm$^2$ 
and $I$ = 5 $\times$ 10$^{14}$ W/cm$^2$), for the frequency-doubled Ti:Sapphire laser: $\lambda=390$nm. We found that the 
linear response in both the TDDFT and SAE models are in good agreement, despite differences in the asymptotic behavior of their 
effective potential. This is reflected in the fundamental frequency in the harmonic spectra. We observed a secondary maxima below 
the classical cutoff frequency and this was found to be independent of the pulse duration. We investigated the effect of changing the 
CEP of the laser pulse on the harmonic yield, and found that the sensitivity of the harmonic spectra to the CEP is apparent at the 
higher intensity for the TDDFT calculation, which is consistent with previous experimental results. Finally, we presented a time-frequency 
representation illustrating the instants of high-frequency ultra-short bursts of light.

Further work will involve optimization of both the TDDFT and SAE effective potentials to give a more accurate and realistic description 
of the system under investigation. It would be of interest to study pulse envelope effects on the HHG process using these models. 

\vspace{1cm}
\begin{acknowledgments}
This work is supported by the U. K. Engineering and Physical Sciences Research Council (EP/C007611/1). 
We are grateful for the support of the University of Bergen in this work.
\end{acknowledgments}

\end{document}